\documentclass[%
superscriptaddress,
preprint,
amsmath,
amssymb, 
aps, 
pre,
floatfix 
]{revtex4-1}

\usepackage{amsfonts}       
\usepackage{amsmath}       
\usepackage{amssymb}
\usepackage{nicefrac} 
\usepackage{graphicx}
\usepackage{dcolumn}
\usepackage{bm}
\usepackage{hyperref}
\usepackage{xcolor}  
\usepackage{array}
\usepackage[export]{adjustbox}
\usepackage{multirow}

\usepackage{changes} 

\usepackage{xifthen} 

\usepackage{todonotes}

\newcommand{\avg}[1]{\left\langle#1\right\rangle} 

\makeatletter
\setremarkmarkup{\todo[color=Changes@Color#1!20,size=\scriptsize]{#1: #2}}
\makeatother

\definechangesauthor[name={Mauricio}, color=red]{M} 
\definechangesauthor[name={Osame}, color=blue]{O} 
\newcommand*{\adM}[2][]{%
    \ifthenelse{\isempty{#1}}{%
        \added[id=M]{#2}%
    }{%
        \added[id=M,remark={#1}]{#2}%
    }%
}
\newcommand*{\deM}[2][]{%
    \ifthenelse{\isempty{#1}}{%
        \deleted[id=M]{#2}%
    }{%
        \deleted[id=M,remark={#1}]{#2}%
    }%
}
\newcommand*{\reM}[3][]{%
    \ifthenelse{\isempty{#1}}{%
        \replaced[id=M]{#2}{ #3}%
    }{%
        \replaced[id=M,remark={#1}]{#2}{ #3}%
    }%
}

\newcommand*{\adO}[2][]{%
    \ifthenelse{\isempty{#1}}{%
        \added[id=O]{#2}%
    }{%
        \added[id=O,remark={#1}]{#2}%
    }%
}
\newcommand*{\deO}[2][]{%
    \ifthenelse{\isempty{#1}}{%
        \deleted[id=O]{#2}%
    }{%
        \deleted[id=O,remark={#1}]{#2}%
    }%
}
\newcommand*{\reO}[3][]{%
    \ifthenelse{\isempty{#1}}{%
        \replaced[id=O]{#2}{ #3}%
    }{%
        \replaced[id=O,remark={#1}]{#2}{ #3}%
    }%
}

\hypersetup{
    unicode=false,          
    pdftoolbar=true,        
    pdfmenubar=true,        
    pdffitwindow=false,     
    pdftitle={SOCBalanced},    
    pdfauthor={Dr. Mauricio Girardi-Schappo},     
    pdfsubject={SOC balanced networks},   
    pdfcreator={Dr. Mauricio Girardi-Schappo},   
    pdfproducer={My computer},  
    pdfkeywords={SOC,} {Balanced} {networks}, 
    pdfnewwindow=true,      
    colorlinks=false,       
    linkcolor=red,          
    citecolor=green,        
    filecolor=magenta,      
    urlcolor=cyan           
}

\begin{document}

\title{Self-organized critical balanced networks: a unified framework}

\author{Mauricio Girardi-Schappo$^*$}
\affiliation{Universidade de S\~ao Paulo, FFCLRP, Departamento de F\'isica, Ribeir\~ao Preto, SP, Brazil}

\author{Ludmila Brochini}
\affiliation{Universidade de S\~ao Paulo, Instituto de Matem\'atica e Estat\'istica, S\~ao Paulo, Brazil}

\author{Ariadne A. Costa}
\affiliation{Universidade Federal de Goi\'as - Regional Jata\'i, Unidade Acad\^emica Especial de Ci\^encias Exatas, Jata\'i, GO, Brazil}

\author{Tawan T. A. Carvalho}
\affiliation{Universidade Federal de Pernambuco,  Departamento de Física, Recife, PE, Brazil}

\author{Osame Kinouchi}
\email{osame@ffclrp.usp.br}
\thanks{girardi.s@gmail.com; These authors contributed equally to this work.}
\affiliation{Universidade de S\~ao Paulo, FFCLRP, Departamento de F\'isica, Ribeir\~ao Preto, SP, Brazil}


\begin{abstract}
Asynchronous irregular (AI) and critical states
are two competing frameworks
proposed to explain spontaneous neuronal activity.
Here, we propose a mean-field model with simple
stochastic neurons that generalizes the 
integrate-and-fire network of Brunel (2000).
We show that the point with balanced
inhibitory/excitatory synaptic weight ratio 
$g_c \approx 4$ corresponds to a second order 
absorbing phase transition usual in self-organized critical 
(SOC) models. At the synaptic balance point $g_c$,
the network exhibits power-law neuronal avalanches
with the usual exponents, whereas for nonzero external field
the system displays the four usual synchronicity states
of balanced networks. We add homeostatic inhibition and
firing rate adaption and obtain a self-organized quasi-critical balanced state
with avalanches and AI-like activity. Our model might explain
why different inhibition levels are obtained in different experimental
conditions and for different regions of the brain, since
at least two dynamical mechanisms are necessary to obtain a truly balanced
state, without which the network may hover in different regions
of the presented theoretical phase diagram.
\end{abstract}

\maketitle

\section*{Introduction}

Spontaneous brain activity happens in the form of
nonlinear waves of action potentials that spread throughout the cortex.
These waves are usually characterized by their
sizes and duration under the critical brain hypothesis
(see~\cite{Hesse2015,Cocchi2017,Wilting2019} for recent reviews),
or by the underlying regularity and global synchronicity
displayed by the firings of neuron
populations under the balanced network 
hypothesis~\cite{Steveninck1997,Brunel2000,Ecker2010}.
These two frameworks are often taken as discordant to each
other~\cite{Wilting2018}, because
a critical activity implies in long-range spatio-temporal
correlations~\cite{Linkenkaer2001,Haimovici2013,Girardi2016},
whereas asynchronous irregular (AI) activity comes from
Poissonian spike trains of interacting neurons~\cite{Softky1993,Brunel2000,Ecker2010}.

Networks working at the critical point have
the advantages of optimizing their dynamic range of
response~\cite{Kinouchi2006,Shew2009},
memory and learning
processes~\cite{Arcangelis2006,delPapa2017},
computational power~\cite{Beggs2012}
and their flexibility to process
information~\cite{Mosqueiro2013}.
However, this highly susceptible state
needs to be achieved and maintained spontaneously by
an autonomous biological
mechanism~\cite{Bonachela2010}, introducing
the problem of self-organized criticality (SOC).
The attempts to model SOC in the context of neuronal networks
involve mainly adaptive
synapses~\cite{Levina2007,Costa2015} and adaptive neuronal
gains~\cite{Kinouchi2019}. Both of these approaches 
are proven to have
theoretically equivalent effects in the system's
phase space trajectory~\cite{Kinouchi2019},
resulting in a self-organized quasi-critical (SOqC) system.
SOqC is a highly sensitive nearly-critical state
able to maintain self-sustained activity, to reproduce
the power-law exponents observed in 
experiments and produce large events (the ``dragon
kings'')~\cite{Costa2015,Costa2017,Kinouchi2019}.
Other authors claim that the power-laws in the brain
are observed because experiments
sample its activity long enough to average out
its spatial disorder into a Griffiths
phase~\cite{Moretti2013,Girardi2016}.

On the other hand, AI activity is intuitively
expected to arise if the resting brain presented 
a statistically fair
random noise. This state is characterized by irregular firing of
individual neurons and global lack of synchronicity,
leading Brunel~\cite{Brunel2000} to conjecture
AI, by visual inspection,
as the typical spontaneous activity of the brain.
This hypothesis is backed by theoretical evidence
that AI states minimize redundancy~\cite{Hyvarinen2000}
and speed-up the processing of inputs~\cite{Van1996,Brunel2000},
whereas excitation/inhibition (E/I) balanced networks may serve to
construct high-dimensional population codes~\cite{Deneve2016}.
This situation is usually modeled by the
so-called balanced networks, in which the system presents
a transition from high to low, or even null,
average activity~\cite{Brunel2000} when the proportion, $g$,
of excitatory to inhibitory synaptic strength is varied.
The transition point, $g_c=4$, is called balanced because
the excitatory population, corresponding to 80\% of the neurons
in the network (a fraction that is
estimated from cortical data for glutamate-activated
synapses~\cite{Somogyi1998}) is balanced by inhibitory
synapses four times as strong as the excitatory ones.

Here, we introduce a model of an E/I network and
show that the balanced point is a critical point
usual for SOC models, unifying both approaches.
Our model predicts that the balance point can be shifted away from
the usual $g_c\approx4$ for regimes in which synaptic couplings
and neuronal gains are relatively small, keeping
fixed the ratio of excitatory neurons in 80\%.
The balance point displays power-law avalanches with exponents
that match experiments~\cite{Beggs2003}. In fact, our model
presents a critical line over the balance parameter
when the average input over the network equals the
average thresholds of the neurons. It also presents all
the synchronicity states found by Brunel~\cite{Brunel2000},
namely the synchronous regular (SR), synchronous irregular (SI),
asynchronous regular (AR), and AI.
We determine the transition lines between these
states. We also show
that firing rate adaptation together with a homeostatic
inhibition mechanism self-organizes the system towards
a quasi-critical state -- a fact that could explain the prevalence of E/I
balanced states in many experiments~\cite{Deneve2016}.
Moreover, the introduced dynamic inhibition leading to the
critical point may explain the enhancing of the dynamic range
observed in experiments when inhibition is considered~\cite{Liu2011},
since the critical state is known to optimize the dynamic range of
excitable systems~\cite{Kinouchi2006}.

There have been attempts to model E/I networks in the context of
criticality~\cite{Poil2012,Lombardi2012,Lombardi2017,DallaPorta2019},
and also connecting criticality to a synchronization
phase transition~\cite{DiSanto2018}.
However, all of these models have limitations that
are naturally solved in our proposed framework. First,
we present our model with all its features in detail, and we finish
the paper comparing our framework to the existing ones.

\section*{The model}

We use discrete-time 
stochastic integrate-and-fire neurons~\cite{Galves2013,Brochini2016}.
A Boolean variable denotes if a neuron fires ($X[t]=1$) or not
($X[t] = 0$) at time $t$. The membrane
potentials evolve (in discrete time) as:
\begin{eqnarray}
V_i^E[t+1]   &=& \left[ \mu V_i^E[t] + I_i^E[t]+ 
\frac{1}{N} \sum_{j=1}^{N_E} W_{ij}^{EE} X_j^E[t]
- \frac{1}{N} \sum_{j=1}^{N_I} W_{ij}^{EI} X_j^I[t] \right]\left(1-X_i^E[t] \right)
 ,\label{VE} \nonumber \\
V_i^I[t+1]   & = & 
\left[\mu V_i^I[t] + I_i^I[t] + \frac{1}{N} \sum_{j=1}^{N_E} 
W_{ij}^{IE} X_j^E[t]
- \frac{1}{N} \sum_{j=1}^{N_I} W_{ij}^{II} X_j^I[t]  \right] \left(1-X_i^I[t]\right), 
\label{VI}
\end{eqnarray}
where $N = N_E + N_I$ is the total number of neurons,
$\mu$ is a leakage parameter and $I_i[t]$ 
are external inputs. 
Here the indices $E$ and $I$ denote the excitatory and 
inhibitory populations. A link $W_{ij}^{EI}$ means a synapse
from an inhibitory neuron to an excitatory neuron (the
second index is always the presynaptic one), and the same
rule applies to $W^{EE}, W^{II}$ and $W^{IE}$.
Notice that if some neuron fires at time $t$, in the next time
step its voltage is reset to zero due to the factor
$(1-X_i[t])$. Also, the inhibitory character of the 
synapses is given by the negative sign in front of the 
summations: the $W^{EI}$ and $W^{II}$ are always absolute
positive values. We notice that our network is a 
complete graph, instead of the sparse network examined
by Brunel~\cite{Brunel2000}. However, 
we will show that there is
no qualitative difference between the two phase diagrams
of both models.

\begin{figure}[b!]
\begin{center}
\centerline{$\begin{array}{ll}
\textnormal{\textsf{\textbf{a}}} & \textnormal{\textsf{\textbf{b}}}\\
\includegraphics[width=0.45\textwidth]{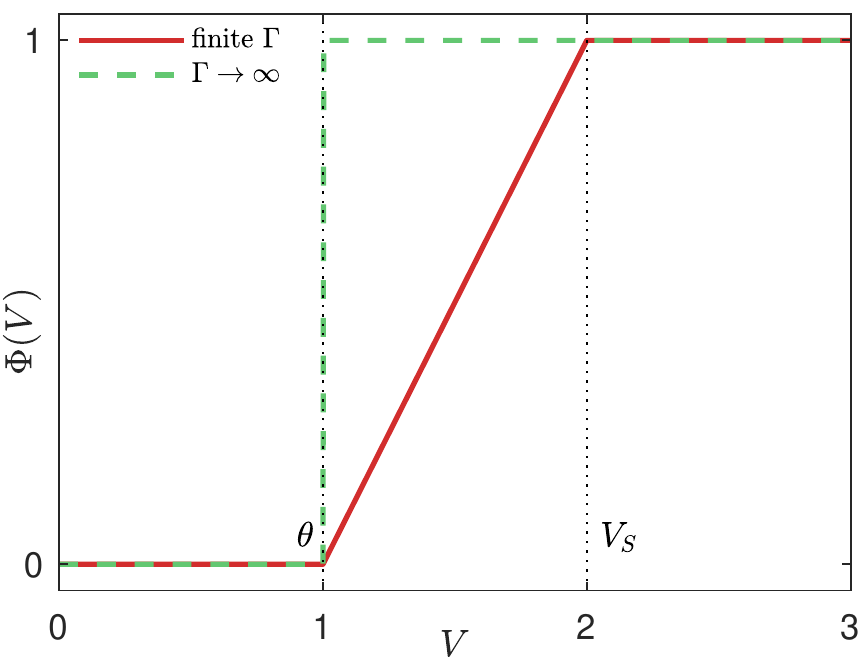} & \includegraphics[width=0.45\textwidth]{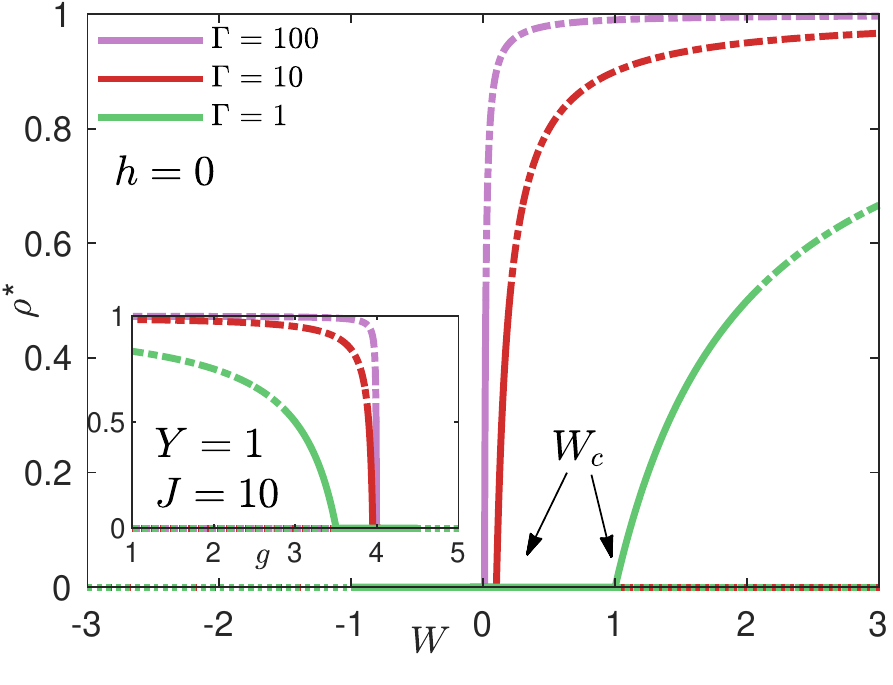}
\end{array}$}
\caption{{\bf Firing probability function $\Phi(V)$ and continuous phase transition.}
{\bf a,} Solid: soft firing threshold $\theta = 1$ with
neuronal gain $\Gamma = 1$, 
so that the saturation voltage is $V_S = 
\theta + 1/\Gamma = 2$.
Dashed: in the $\Gamma \rightarrow \infty$ limit, we recover
the deterministic leaky integrate-and-fire (LIF) model with a
sharp threshold $\theta = V_S = 1$.
{\bf b,} Phase transition for $\rho^* = 
\rho_E = \rho_I$ 
as a function of $W$ along the line $h = 0$, 
for $\Gamma = 1, 10$ and $100$ and $J = 10$
($W_c = 1/\Gamma$).
{\bf b (inset),} Phase transition for $\rho^* $ 
as a function of the ratio $g$ along 
the line $Y = 1$ for the same $\Gamma$ and $J$. The order
parameter $\rho^*$ is equivalent to the firing
frequency $\nu_0$ of Brunel model~\cite{Brunel2000}.
Notice that the FP with $\rho^*>1/2$ 
(dot-dashed) represents cycle-2
due to the refractory period of one time step (Brunel's Synchronous
Regular SR state).}
\label{PHI}
\end{center}
\end{figure}

The individual neurons fire following a piecewise linear  
probability function (see Fig.~\ref{PHI}a):
\begin{equation}
P\left(X =1 \:|\: V\right) \equiv \Phi(V) 
= \Gamma (V-\theta) \: \Theta(V-\theta) \:
\Theta(V_S - V)   
+ \Theta(V-V_S)\:,   \label{eqPHI}
\end{equation}
where $\Gamma$ is the neuronal gain, 
$\theta$ is a firing threshold, $V_S = \theta + 1/\Gamma$
is the saturation potential and
$\Theta(x)$ is the step Heaviside function.
The fact that $0 < \Phi(V) < 1$ in the interval $[\theta,V_S]$
means that the spikes are stochastic, a feature that
intends to model the effects of membrane noises.
Notice that the limit $\Gamma \rightarrow \infty$ reproduces
the (discrete time) deterministic integrate-and-fire neuron
with hard threshold $V_S=\theta$.

\section*{Results}

First, we show that the balance point is a critical point, characterizing its critical
exponents and its avalanche distributions. We then determine the full
phase diagram, locating the synchronicity states (SR, AR, AI, SI). And finally, we
introduce an adaptive dynamics in the inhibitory synapses together with firing
rate adaptation and show that the system then hovers towards the
critical balanced point.

\subsection*{The mean-field limit}

In the Methods section, we derive an 
expression for the mean firing rate of the
excitatory and inhibitory neuronal populations to be used here.
We approximate
the excitatory/inhibitory synaptic weights by their average values, 
$W^{EI} =\avg{W_{ij}^{EI}}$ (for all the E/I combinations),
and write the firing densities
(the fraction of active sites) $\rho_E = 1/N_E \sum_j X_j^E$ and
$\rho_I = 1/N_I \sum_j X_j^I$.
We write the fractions of excitatory and inhibitory
neurons as $p = N_E/N$ and $q = 1-p = N_I/N$, respectively. 
We also consider only the case with a stationary
average input $I = \avg{I_i[t]}$, so that the neurons' membrane
voltages evolve as:
\begin{eqnarray}
V_i^E[t+1]   & = & \left[ \mu V_i^E[t] + I +
 p  W^{EE} \rho_E[t]
- q  W^{EI} \rho_I[t]  \right] \left( 1 - X_i^E[t]\right)\:,
\nonumber\\
V_i^I[t+1]  & = & \left[ \mu V_i^I[t] + I +
p  W^{IE} \rho_E[t]
- q  W^{II} \rho_I[t] \right] \left( 1 - X_i^I[t]\right) \:.
\label{Vs}
\end{eqnarray}
The densities $\rho_E$ and $\rho_I$ of firing neurons
are our order parameters
(similar to the firing frequency $\nu_0$ of the
Brunel model~\cite{Brunel2000}) and 
the synaptic weights $W^{EE}, W^{EI}, W^{IE}$ and
$W^{II}$ are our control parameters. Together with 
$p$, $I$, $\mu$, $\theta$ and $\Gamma$, we have a nine-dimensional
parameter space that will be further reduced.

Analytical results are exact for $\mu=0$.
The case $0<\mu<1$ admits numerical results for any 
leakage $\mu$~\cite{Brochini2016,Costa2017}, but also has analytical solutions
close to the phase transition. For $\rho\geq1/2$, solutions
are independent of $\mu$.
Here, we consider the case $\mu=0$ since it captures
all the important dynamics of the model and the effect of non-zero $\mu$
is described in the Methods.
Making the E/I synaptic 
strengths uniform (Brunel's model A~\cite{Brunel2000}),
we can study the stationary states in terms
of the synaptic balance parameter $g$:
we define $W^{EE} = W^{IE} = J$, 
$W^{II} = W^{EI} =  g J$, and
the weighted synaptic weight, $W =  p J - q g J $.
This choice gives us
a uniform solution (see Methods)
$\rho[t] = \rho_E[t] = \rho_I[t]$ that evolves as:
\begin{equation}
\label{rhosfinal}
\rho[t+1] = \Gamma\:\left(1-\rho[t]\right)\:\left( W \rho[t]+ I  - \theta \right)
\Theta(W \rho[t]+ I  - \theta)\:.
\end{equation}
Its fixed points are given by:
\begin{equation}
\Gamma W \rho^2 + \left(1 + \Gamma h - \Gamma W
\right) \rho - \Gamma h = 0 \:,  \label{eqrhomu0}
\end{equation}
where we defined the supra-threshold external current, $h=I-\theta$,
equivalent to an external field in usual SOC models.

\begin{figure}[t!]
\centerline{$\begin{array}{ll}
\textnormal{\textsf{\textbf{a}}} & \textnormal{\textsf{\textbf{b}}}\\
\includegraphics[width=0.45\textwidth]{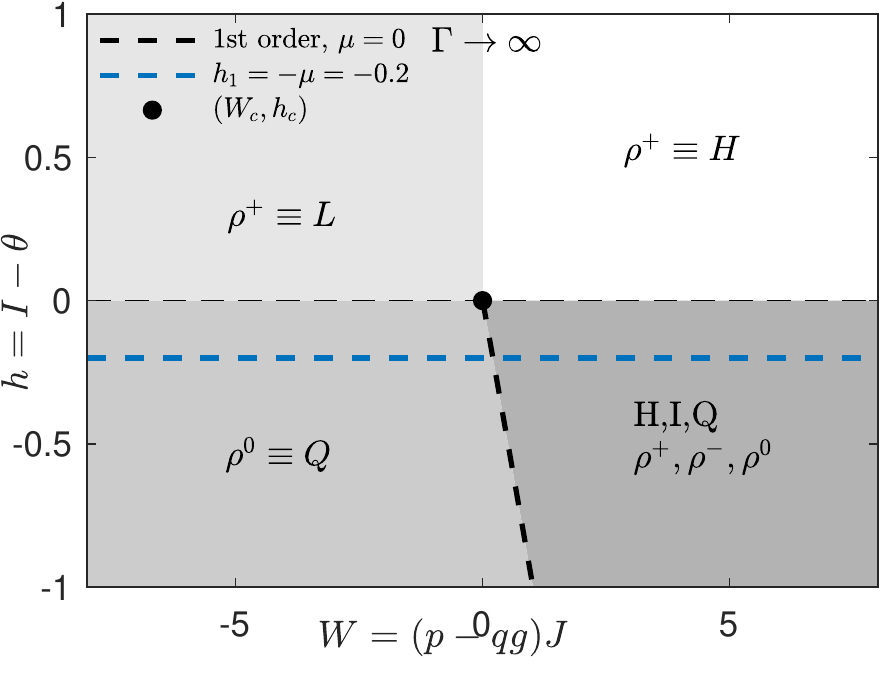} & \includegraphics[width=0.45\textwidth]{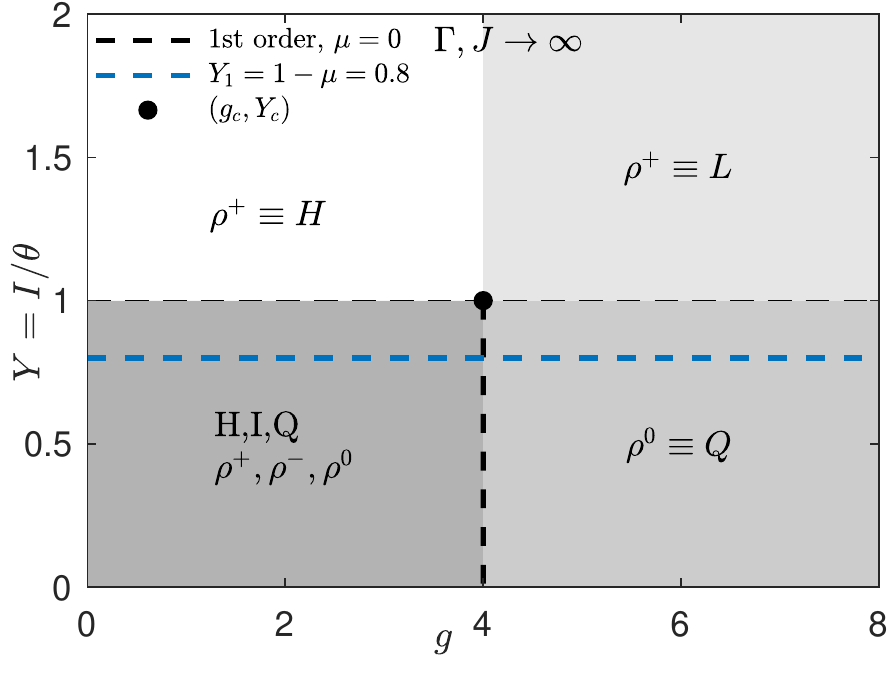}\\
\textnormal{\textsf{\textbf{c}}} & \textnormal{\textsf{\textbf{d}}}\\
\includegraphics[width=0.45\textwidth]{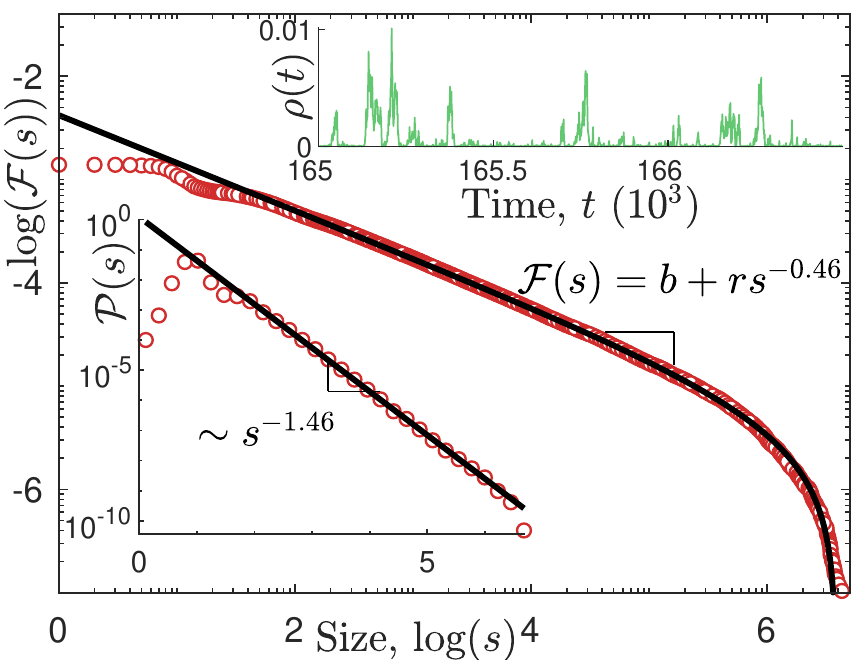} & \includegraphics[width=0.45\textwidth]{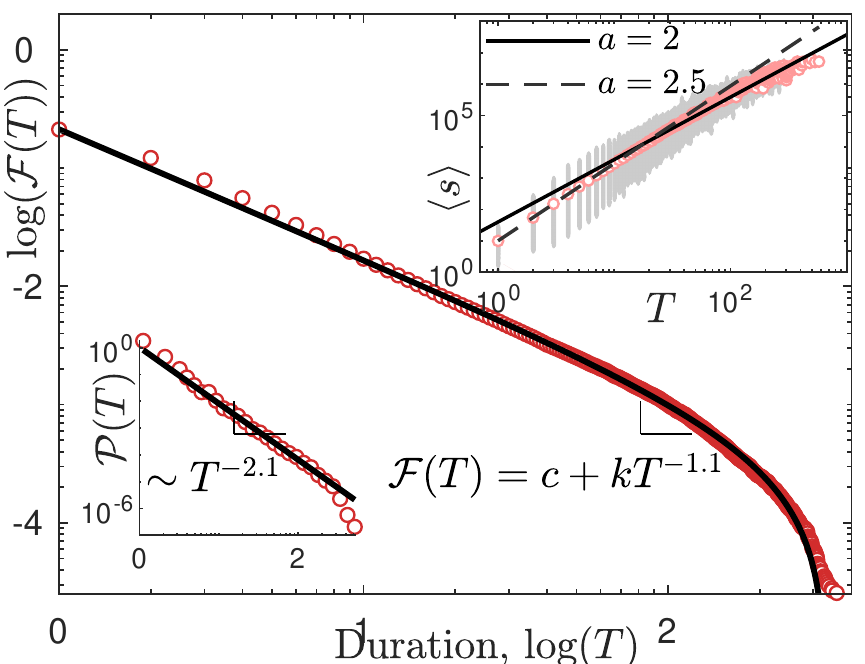}
\end{array}$}
\caption{{\bf Phase diagrams for large $\Gamma J$ and critical avalanches
in the balanced point of the static model.}
{\bf a,} Phase diagram in the $(W,h)$ plane. Notice that the Low (L) and 
High (H) behavior corresponds to a 
single phase $\rho^+$ with different amplitudes.
{\bf b,} Phase diagram in the balanced notation, 
$(g,Y)$ plane. The critical point
lies at $(W_c = 0, h_c = 0)$ or 
$(g_c = 4, Y_c = 1)$ [bullet, equation~\eqref{gc}].
The Q phase loses stability at the horizontal dashed
line $Y_c =1-\mu$ 
(or $h_c = -\mu$); we show $\mu=0$ (black and thin dashed line) and
$\mu = 0.2$ (blue dashed line).
This diagram should be compared to Brunel's Fig.~1A~\cite{Brunel2000}.
{\bf c,} Time series of network activity density (top inset),
distribution (bottom inset) and complementary CDF (main plot)
of avalanche sizes in the static model, yielding $\tau=1.46$ (solid line).
{\bf d,} Avalanche size and duration scaling law yields two exponents, $a=2.5$
for small avalanches and $a=2$ for the rest of the data (top inset),
distribution (bottom inset) and complementary CDF (main plot)
of avalanche duration in the static model, yielding $\tau_t=2.1$ (solid line).
Note the deviation of the distributions for small and short avalanches, causing
the cross-over effect in the $\avg{s}\sim T^a$ scaling law.}
\label{rhoxW}
\end{figure}

In order to put Brunel's model~\cite{Brunel2000} into the
standard SOC framework given in the parameter space $(W,h)$
(Fig.~\ref{rhoxW}a),
we introduce the parameter $Y=I/\theta$ -- the fractional external
current -- entirely equivalent to Brunel's input ratio 
$\nu_{\rm ext}/\nu_{\rm thr}$ where 
$\nu_{\rm ext}$ is the average of a noisy
input current and $\nu_{\rm thr}$ is proportional to $ \theta$.
We assume $\theta=1$ without loss of generality.
Also, recall that
$g = p/q - W/(qJ)$, enabling the study of the system in the 
\textit{balanced notation} phase diagram $(g,Y)$
(Fig.~\ref{rhoxW}b).
At this point we have a six
dimensional parameter space $\{\Gamma,J,g,Y,\mu,p\}$,
but we fix $p=0.8$, and hence $q=0.2$,
from cortical data~\cite{Somogyi1998}.
We henceforth call this the static model. If any of the parameters
presented in this section is allowed to vary with time,
then we have the dynamic model that will be
discussed later in this manuscript.

\subsection*{Balanced point as a second order critical point}

Considering the
case $h = I - \theta = 0\equiv h_c$ in equation~\eqref{eqrhomu0},
we obtain an absorbing state
$\rho^0 = 0$  (the quiescent phase, Q) which is
stable for $W<W_c \equiv 1/\Gamma$, 
and an active solution (also called H or L for high or low activity,
respectively):
\begin{equation}  \label{trans}
\rho^*  = \frac{\Gamma W-1}{\Gamma W} = \frac{W - W_c}{W}\:,
\end{equation}
stable for $W > W_c$. 
This is a transcritical bifurcation 
usual in SOC models, see Fig.~\ref{PHI}b.
Equation~\eqref{trans} is exact for $\mu = 0$. For $\mu > 0$,
$h_c=-\mu$, and a good
approximation near the phase transition results in
$W_c = (1-\mu)/\Gamma$, such that
\begin{equation}
\rho^* = \frac{W - W_c}{W}(1-\mu)\:.
\end{equation}

For $W\rightarrow W_c$, the active solution reduces to
$\rho^*\sim (W-W_c)^{\beta}$, with $\beta=1$ being
the critical exponent associated with the coupling 
intensity. Analogously, we
can isolate $h$ from equation~\eqref{eqrhomu0} and 
expand for small $\rho$ (due to small external $h$)
on the critical point $W=W_c$ to obtain $\rho^*\sim
[h/W_c]^{1/\delta_h}$ with $\delta_h=2$
the critical exponent associated with the external field.
The susceptibility, $\chi$, exponent is obtained by taking the derivative of $\rho$
with respect to $h$ at $h=0$ in equation~\eqref{eqrhomu0},
using $\Gamma=1/W_c$, and expanding for $W\rightarrow W_c$.
This procedure results in $\chi\sim|W-W_c|^{-\gamma'}$ with $\gamma'=1$.
These mean field exponents are
compatible with the directed percolation (DP)
universality class~\cite{Dickman1999}, 
the framework proposed to govern
SOC systems~\cite{Dickman1998,Brochini2016,
Costa2017,Kinouchi2019}.
In the DP, the variance of the network activity defines
the exponent $\gamma$ by ${\rm Var}(\rho)\sim|W-W_c|^{-\gamma}$
with $\gamma=0$~\cite{Dickman1999}. This explains the jump in the
coefficient of variation of the network activity
observed by Brunel~\cite{Brunel2000}, since a zero-valued $\gamma$
exponent indicates a discontinuous jump in
the variance of $\rho$.

In the balanced notation, for the external input
$Y_c=h_c+1=1$, and using the relationship between $W$ and $g$,
we can write the critical point as:
\begin{equation}
\label{gc}
g_c = p/q - \dfrac{1}{q \Gamma J} = 
4 - \dfrac{1}{0.2 \Gamma J} \:,
\end{equation}
where we chose the usual
fractions $p = 0.8$ and $q = 0.2$ for cortical
neurons.
Our result generalizes the known condition $g_c\approx 4$:
If the term $1/(q \Gamma J)$ turns out significant, 
the transition shifts towards lower values of $g$
(see the inset of Fig.~\ref{PHI}b). This means that when
the synaptic strengths, $J$, or the firing rate gains, $\Gamma$, are small,
the network will not need much inhibition, hence the smaller $g$ value.
This was only observed numerically
in~\cite{Brunel2000}.
The large $\Gamma J$ regime, which approximates the LIF neurons
in equation~\eqref{eqPHI}, recovers the usual balanced condition,
$g_c\approx 4$, obtained by Brunel~\cite{Brunel2000}. 

Rewriting equation~\eqref{trans} using $g_c$, we get:
\begin{equation}
    \rho^* = \dfrac{pJ - q g J - 1/\Gamma}{pJ - q g J} 
     = \dfrac{g_c - g}{p/q-g},
\end{equation}
and we recover the phase transition for $g\rightarrow g_c$ with
$\beta=1$ and $\gamma'=1$.
Nevertheless, the states are flipped in the $g$-axis: the active
state now happens for $g<g_c$. The expansion for $Y\gtrsim Y_c\equiv1$ 
at $g=g_c$ also yields $\delta_h=2$. The same
holds for $\mu>0$, but with $Y_c=1-\mu$.
So, balanced networks share the same 
second order phase transition of SOC models.

\subsection*{Avalanches in the balanced critical point}

The balanced point (which is now proven to be a DP critical point)
displays avalanche-like activity
that can be measured from the $\rho[t]$ time series (see the top
inset of Fig.~\ref{rhoxW}c).
We simulate the static model as a mean-field network of
$N=10^6$ neurons
with $\Gamma=1$, $J=10$, $g=g_c=3.5$, $Y=Y_c=1$, $\mu=0$
and $p=0.8$.
We define an avalanche as the sum of all the spikes between
two consecutive silent absorbing states~\cite{Girardi2018}.
Avalanches are sparked independently
when the activity goes to zero.
We fit cutoff power laws to the complementary cumulative distributions (CDF)
of avalanche sizes, $\mathcal{F}(s)\equiv\mathcal{P}(S>s)=b+rs^{-\tau+1}$, and duration,
$\mathcal{F}(T)\equiv\mathcal{P}(T'>T)=c+kT^{-\tau_t+1}$
($b$, $r$, $c$ and $k$ are fit constants related to
the cutoff of the power laws~\cite{Girardi2013b}).
This representation provides a clearer
visualization of the data, because it is a continuous function of
its variables; it also has very reduced noise, its precision does not
depend on the size of the bins of the distribution's histogram,
and it has a better defined cutoff~\cite{Girardi2013b}.

Least squares fit yields
$\tau=1.46(4)$ and $\tau_t=2.1(1)$ for avalanche sizes and duration exponents,
respectively (main plots of Figs.~\ref{rhoxW}c and 2d).
We also fit the histogram of sizes, $\mathcal{P}(s)\sim s^{-\tau}$,
and duration, $\mathcal{P}(T)\sim T^{-\tau_t}$,
and obtain the same values for $\tau$ and $\tau_t$
(bottom insets of Figs.~\ref{rhoxW}c and 2d).
The expected values for these exponents in the DP universality class
are $\tau_{\rm DP}=1.5$ and $\tau_{t,{\rm DP}}=2$, but the finite-size
effects cause the deviation of the distributions for small $s$ and short $T$,
biasing the fitted values of the exponents. Nevertheless,
both the fitted values and the theory values match experiments~\cite{Beggs2003}.

The average avalanche size scales with avalanche duration
according to $\avg{s}\sim T^a$ such that
$a=(\tau_t-1)/(\tau-1)$~\cite{Dickman1999,Girardi2016b,Girardi2018}.
The theoretical value of $a$ in DP is $a_{DP}=2$. In our simulation,
we observe a cross-over in the $\avg{s}$ \textit{vs.} $T$ data
(top inset of Fig.~\ref{rhoxW}d),
such that for large and long avalanches $a=2$ (exactly equal to the
theory),
and for small and short avalanches, $a=2.5$ due to the finite-size
deviations in the $\mathcal{P}(s)$ and $\mathcal{P}(T)$
distributions. The fitted values $\tau=1.46$ and $\tau_t=2.1$
result in $a_{\rm fit}=2.4$, which is between the cross-over
exponents, hence agreeing with the $\avg{s}$ \textit{vs.} $T$ data.

\subsection*{The synaptic balance \textit{vs.} external current phase diagram}

\begin{figure}[t!]
\centerline{$\begin{array}{ll}
\textnormal{\textsf{\textbf{a}}} & \textnormal{\textsf{\textbf{b}}}\\
\includegraphics[width=0.45\textwidth]{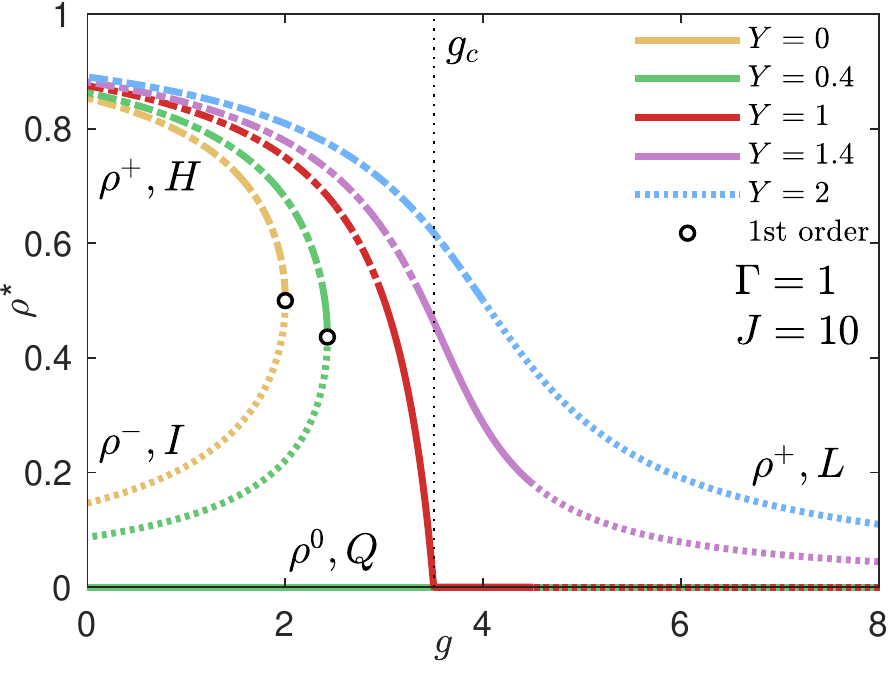} & \includegraphics[width=0.45\textwidth]{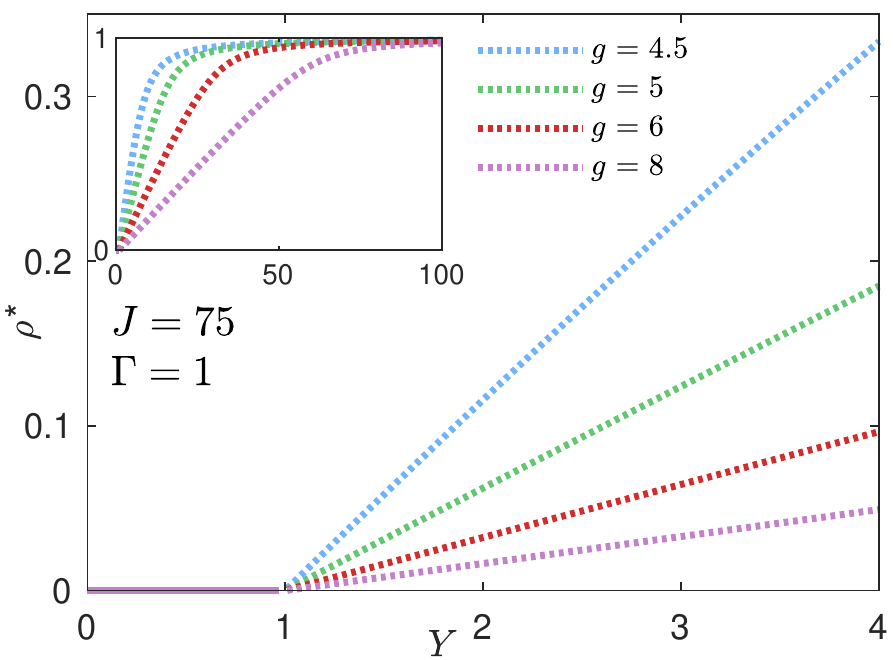}\\
\textnormal{\textsf{\textbf{c}}} & \textnormal{\textsf{\textbf{d}}}\\
\includegraphics[width=0.45\textwidth]{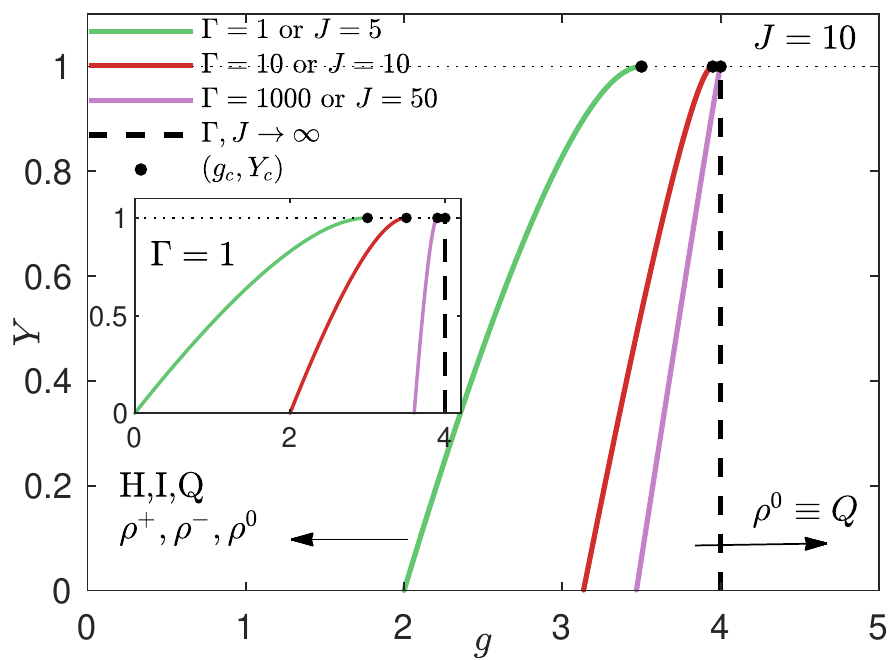} & \includegraphics[width=0.45\textwidth]{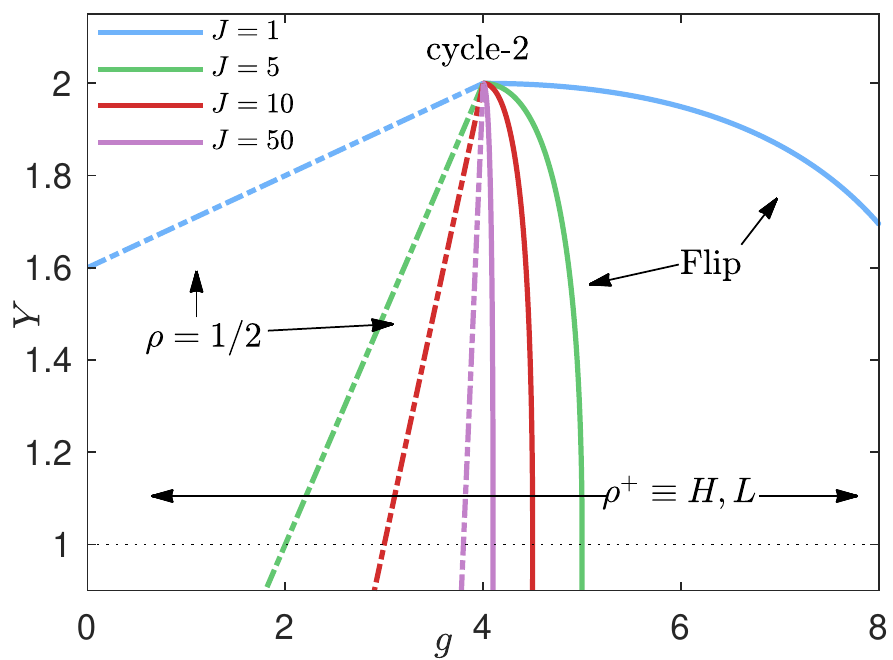}
\end{array}$}
\caption{\label{rhoxg}{\bf Firing density $\rho$ \textit{vs.} $g$
and $Y$ and phase diagrams in the plane $(g,Y)$.} 
{\bf a,} Firing density as a function of $g$ for many $Y$
[equation~\eqref{FOT2}], highlighting the activity states
H for high, L for low (both corresponding to $\rho^+$),
I for the intermediary and unstable $\rho^-$ (originating in
a fold bifurcation, or first order phase transition)
and Q for $\rho=0$.
{\bf b,} Firing density as a function of $Y$ for many $g$;
linear behavior occurs for small $Y$, but saturates for large
external current (inset).
This plot has no SR behaviour and should be compared to Fig.~1B
(right) in Ref.~\cite{Brunel2000}.
The dotted lines for $\rho^+$ in both panels \textbf{a} and
\textbf{b} correspond to marginally stable fixed points.
{\bf c,} First order phase transition (fold bifurcation) lines
[equation~\eqref{foldbif}] for $J=10$ and many
neuronal gains $\Gamma$; below the line $Y=1$, we have
the Q state, whereas to the left of each $Y_1$ line we also have
the states H and I.
{\bf c (inset),} Same diagram as in the main plot, but
for $\Gamma=1$ and many synaptic intensities $J$.
{\bf d,} Bifurcation lines for $\Gamma=1$ towards cycle-2 at $\rho^+=1/2$,
equation~\eqref{srbif} (dot-dash line), and flip [equation~\eqref{YF},
solid line].
SR occurs above each $\rho^+=1/2$ line and SI occurs above each
flip line.}
\end{figure}

The solutions to equation~(\ref{eqrhomu0}) 
for $h \neq 0$ (or $Y \neq 1$) are:
\begin{equation}
\rho^\pm  =  \dfrac{\Gamma W -\Gamma h -1
\pm\sqrt{(\Gamma W-\Gamma h-1)^2 
+ 4 \Gamma^2 W  h}}{2\Gamma 
W} \:, \label{FOT} 
\end{equation}
The balanced notation can be recovered
letting $h = Y - 1$ (recall that $\theta = 1$) and
$W=pJ(1-g\gamma)$, where $\gamma = q/p$ (not to be confused with
the critical exponent for the variance of $\rho$):
\begin{eqnarray}
\rho^\pm & = & \dfrac{1 - g \gamma - (Y - 1)/(pJ) - 
1/(p\Gamma  J) \pm \sqrt{\Delta}}{2 (1 - g\gamma)}\:, \label{FOT2} \\
\Delta & \equiv & \left[1 - g\gamma - \dfrac{Y-1}{pJ} - 
\dfrac{1}{p\Gamma J}\right]^2 + 4 \left(1-g\gamma\right) \dfrac{Y-1}{pJ}
\nonumber \:.
\end{eqnarray}
Notice that there is no divergence when 
$g = p/q$ ($W=0$) because equation~(\ref{eqrhomu0})
then gives $\rho = \Gamma h/(1+\Gamma h)$. Solutions
for equation~\eqref{FOT2} are plotted in Figs.~\ref{rhoxg}a
and~\ref{rhoxg}b for different values of $g$ and $Y$.
Equation~\eqref{FOT2} yields the activity states of
high and low activity (H and L, the $\rho^+$),
and also the intermediary state (I, the unstable $\rho^-$ branch).
However, there is also a stable quiescent solution Q, $\rho^0=0$,
for $Y \leq Y_c$ due to the step function
in equation~\eqref{rhosfinal}; but Q is 
unstable over the critical line for $Y_c=1-\mu$ 
if $g < g_c$ due to the transcritical bifurcation
at the critical point (horizontal dashed lines in Fig.~\ref{rhoxW}b).

For $Y < Y_c$, there is a first order
transition line where the $\rho^+$ branch appears and start
to coexist with the stable $\rho^0$ solution,
generating the intermediary unstable state $\rho^-$
through a fold bifurcation (Fig.~\ref{rhoxg}a).
The first order transition line, shown in Fig.~\ref{rhoxg}c, is 
(see Methods):
\begin{eqnarray}
Y_1(g) = 1 - \dfrac{1}{\Gamma} \left[
\sqrt{p\Gamma J (1 - g\gamma)} - 1 \right]^2
\label{foldbif}\:.
\end{eqnarray}
This transition depends on $J$ and $\Gamma$
and it ends at $Y_1(g_c)=Y_c=1$.
For large $\Gamma J$, we recover Brunel's~\cite{Brunel2000} diagram,
with an almost vertical $Y_1$ line at $g_c$, as shown in
Fig.~\ref{rhoxW}b.

For $Y > Y_c$, the $\rho^-$ (I) solution disappears and
the Q phase is unstable.
The $\rho^+$ solution varies continuously from H to L activity state.
There is no phase transition by changing $g$,
although a clear change of behavior (large derivative) 
occurs close to $g_c$.
This happens because,
for $g < g_c$, we have a supercritical 
phase with self-sustained
activity being fueled by the external input (the H state)
whereas for the $g > g_c$ region, there is
no self-sustained activity (Fig.~\ref{rhoxg}a),
making the L state appear due entirely
to the external current.
Nevertheless, both the H and L states bifurcate (see Fig.~\ref{rhoxg}d)
giving rise to distinct oscillatory behaviors to be discussed in
the next section.

Even though equation~\eqref{FOT2} admits H solutions that have
$\rho^+>1/2$ as a stable fixed point, the original system
does not. This happens because the neurons defined in equation~\eqref{Vs}
have a refractory period of 1 time step (since $\Phi(0)\equiv0$), yielding
a maximum firing rate of $\rho=1/2$. Thus, mean-field solutions
that have $\rho^+>1/2$ correspond to a cycle-2 dynamics of the form $\rho[t+1] = 1/2 - \rho[t]$
for the network.
The amplitude of this cycle depends on the initial condition,
so $\rho>1/2$ is marginally stable.
The transition line towards this cycle-2 dynamics is
given by (see Methods):
\begin{equation}
\label{srbif}
Y_{SR}(g) = \dfrac{p J}{2} (1-g\gamma) + 1 
+\dfrac{1}{\Gamma}\:.
\end{equation}
This curve is plotted in Fig.~\ref{rhoxg}d for different values of $J$.
When $\Gamma J$ is large, it becomes almost vertical, turning
almost all the H state into cycle-2 activity for $g<g_c$.
This transition is not a standard bifurcation of maps, but rather
a particularity of the model raised by the refractory period.

The L state bifurcates through a flip. The consequence is that the $\rho[t]$
map given in equation~\eqref{rhosfinal} starts to cycle
between negative and positive values,
even though the negative ones are turned to $\rho=0$ due to
the step function.
The bifurcation line is given by (see Methods):
\begin{equation}
Y_{F}(g) =  1 - \dfrac{1}{\Gamma}+ pJ(g\gamma-1) 
        +\dfrac{2}{\Gamma}\sqrt{1-
        p\Gamma J(g\gamma -1)} \:.
       \label{YF}
\end{equation}
This curve is shown in Fig.~\ref{rhoxg}d for different values of $J$.
It becomes almost vertical for large $\Gamma J$, turning
almost all the L state into
oscillations for $g>g_c$.

\subsection*{Spiking patterns emerging for external input currents}

The H and L activity states (both given by $\rho^+$) can be further split into
four different states, depending on the global synchronicity and local regularity
of the firing patterns of the network.
The network may or may not be globally synchronized; and the synchronization
(or the lack of it) may happen in a locally regular or
irregular fashion (meaning that each individual neuron either
spikes regularly or not, respectively)~\cite{Brunel2000}. Thus, we
have the four different oscillatory states: globally synchronous and locally regular
(SR); globally asynchronous and locally regular (AR); asynchronous and irregular
(AI); and synchronous irregular (SI).

\begin{figure}[t!]
\begin{center}
\centerline{$\begin{array}{@{}l@{}l@{}}
\textnormal{\textsf{\textbf{a}}}& \textnormal{\textsf{\textbf{b}}}\\
\includegraphics[width=0.4\textwidth,valign=t]{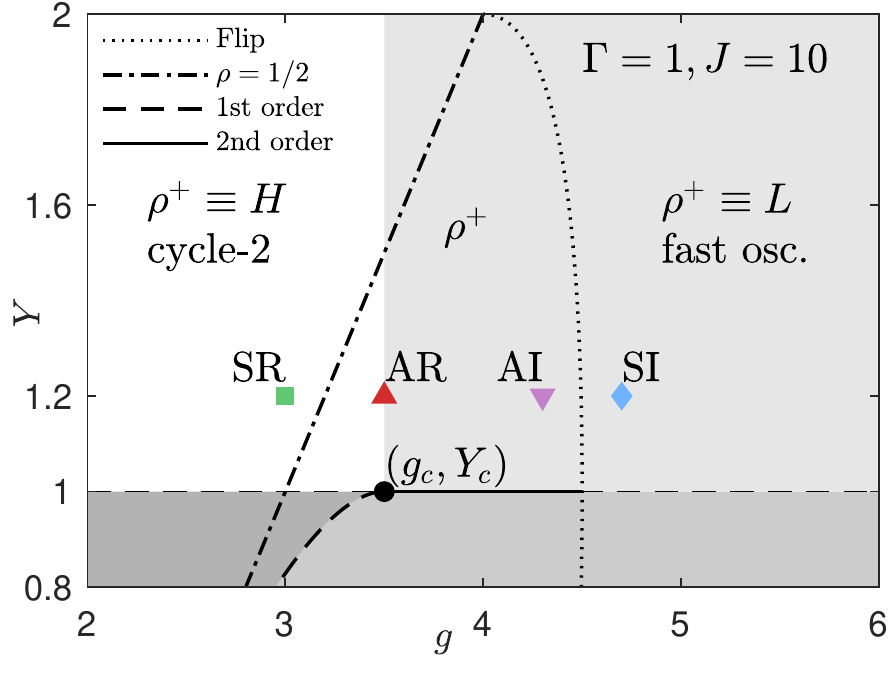} & \includegraphics[width=0.57\textwidth,valign=t]{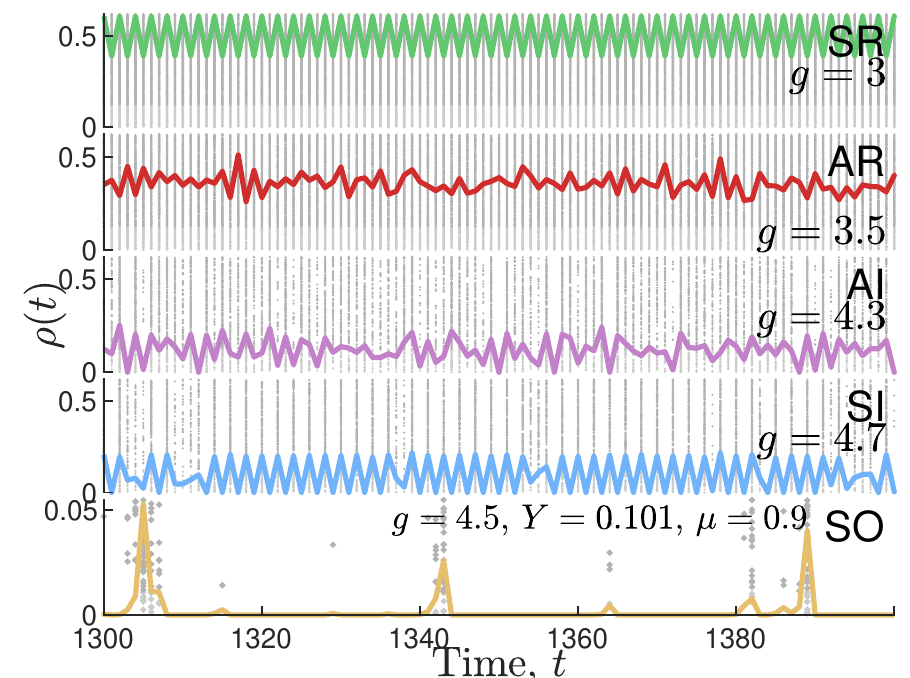}
\end{array}$}
\caption{\label{PD}{\bf Spiking patterns.}
{\bf a,} Full phase diagram of our model for $\mu=0$, $\Gamma=1$, $J=10$, $Y=1.2$, and finite $\Gamma J$;
from left to right we have the synchronous regular (SR/cycle-2, $g=3$), 
asynchronous regular (AR/$\rho^+$/High, $g=3.5$), asynchronous irregular
(AI/$\rho^+$/Low, $g=4.3$) and synchronous irregular (SI/fast oscillations, $g=4.7$).
{\bf b,} Network simulation results, showing the firing density time series, $\rho[t]$,
and the raster plot of 1000 neurons for the cases highlighted in panel~\textbf{a}.
{\bf b (bottom),} Example of slow oscillations (SO).
They appear as sparse avalanches close to $Y=1-\mu$. Parameters are the same
as in panel~\textbf{a}, except for $\mu=0.9$ and $Y=0.101$.
}
\end{center}
\end{figure}

Fig.~\ref{PD}a shows the location of each of these four states in the
complete phase diagram of our model for finite $\Gamma J$.
The main feature defining the synchronicity of the SR and SI states is that
the frequency of the firing density oscillations is well-defined,
meaning that the network is indeed operating as a global oscillator.
The firing density of AR and AI states, on the other hand, behaves mostly as noise.

The mean-field solutions to equation~\eqref{FOT2} give us
only the global state of synchrony. To probe for the microscopic regularity
in spiking patterns, we simulated equation~\eqref{Vs} for a network
of $N=10^6$ neurons, $J=10$, $\Gamma=1$, $p=0.8$, $\mu=0$, $Y=1.2$ and
four values of $g$:
$g=3.0$ (excitation-dominated network), $g=3.5$ (weakly excitation-dominated network),
$g=4.3$ (weakly inhibition-dominated network), and $g=4.7$ (inhibition dominated
network). To obtain slow oscillations, we simulated a large leakage regime,
$\mu=0.9$, with a small input current, $I=0.001$, giving $Y=0.101$
(recalling that the critical line is at $Y_c=1-\mu=0.1$),
in the inhibition-dominated region of the diagram, $g=4.5$.
The $(g,Y)$ points for $\mu=0$ are marked in Fig.~\ref{PD}a, and the results
of the simulations are shown Fig.~\ref{PD}b.

The self-sustained activity regime, when summed up to the
external current, gives rise to regular microscopic behavior, SR and AR.
The AR state turns into SR
as excitation is increased when the system undergoes the $\rho=1/2$ transition
(see Fig.~\ref{PD}a).
Note that this
is not a standard bifurcation for map systems, but rather a particular
feature of our model due to its 1~ms refractory period,
causing the marginally stable cycle-2 activity.

The addition of an external current to the quiescent regime generates
irregular microscopic activity. This happens because interactions
are dominated by inhibitory synapses, such that the negative currents
transmitted through the network compensate for the average external positive input.
The AI state turns into SI as inhibition is increased when the system undergoes
a flip bifurcation (see Fig.~\ref{PD}a). From then on, the system is so
saturated with inhibition that the activity often dies out just to be
sparked again by the constant input, generating fast oscillations
that, globally, look almost like cycle-2.

Differently from Brunel~\cite{Brunel2000}, all our model's randomness
is condensed in the stochastic firing of individual neurons. The network
is topologically regular (a full graph),
and the synaptic and external inputs are homogeneous parameters.
Thus, the only requirements to generate these four behaviors
are: an intrinsic noise source, and a refractory period that is equal to or larger
than the synaptic transmission time (which is 1~ms in our model).

When leakage is introduced into the model through $\mu>0$,
the voltage of the neurons become distributed in an exponential
way due to the slow recovery of the membrane potential
after the spike~\cite{Brochini2016}. This generates a chain of delayed firings intercalated
by long periods of silence that depend on $\mu$.
The firings form sparse relatively large avalanches
intertwined with small avalanches,
and Brunel~\cite{Brunel2000} called them as slow oscillations (SO,
bottom panel of Fig.~\ref{PD}).
This behavior occurs for external input $Y$ very close to the critical line
$Y\gtrsim Y_c=1-\mu$. Large inputs saturate the activity either
through regular or irregular states, depending on the value of $g$.

\subsection*{Self-organized critical balanced networks}

For the brain to reach and maintain the balanced
state with $g=g_c$ and $Y=Y_c$ without fine tuning,
there has to exist a self-organizing mechanism.
While inhibition frequently increases together with excitation
after the stimulation of a neuron, the reverse
does not seem to happen; that is, excitation does not compensate for
inhibition when the neuron is suppressed~\cite{Deneve2016}. 
To model this feature, we introduce an adaptive mechanism
on the weights of the inhibitory synapses
inspired on the Levina-Herrmann-Geisel (LHG)
dynamics~\cite{Levina2007},
\begin{eqnarray}\label{WII}
W^{II}_{ij}[t+1]  &=&   W^{II}_{ij}[t] + \dfrac{1}{\tau_W} 
(A-W^{II}_{ij}[t]) + u_W W^{II}_{ij}[t] X_j[t] \:,\\
W^{EI}_{ij}[t+1]  &=&   W^{EI}_{ij}[t] + \dfrac{1}{\tau_W} 
(A-W^{EI}_{ij}[t]) + u_W W^{EI}_{ij}[t] X_j[t] \:.
\end{eqnarray}
Here, $\tau_W$ is a (large) recovery time, $A$ is a recovery level and 
$u_W$ is the fraction of the synaptic strength facilitated
when a presynaptic neuron fires.
This mechanism potentiates inhibition by a factor $u_W$
when the presynaptic neuron fires and then
relaxes with a time scale given by $\tau_W$.
The dynamics in the $W_{ij}$ generates a response
in the $g$ axis of the phase diagram through
$g_{ij}^{EI/II}[t] = W_{ij}^{EI/II}[t]/J$.

However, our model has two degrees of freedom, requiring another independent
mechanism to reach the balanced condition autonomously.
This is achieved through firing rate adaptation~\cite{Benda2003},
which we model by a homeostatic dynamics in the firing threshold of the
neurons:
\begin{equation}
        \theta_i[t+1] =\theta_i[t] - \dfrac{1}{\tau_\theta}
\theta_i[t]   + u_\theta \theta_i[t] X_i[t]\:,
\end{equation}
where the parameter $u_{\theta}$ is the fractional increase in the
neuron threshold after it fires, and $\tau_{\theta}$ is a recovery
time scale towards a null theta relative to the external current.
This dynamics is sufficient to drive the system
along the $Y$ axis, because $Y_i[t] = I/\theta_i[t]$.
When the neuron fires, its threshold 
adapts to prevent new firings, otherwise it decays with a large time scale
$\tau_\theta$.

\begin{figure}[t!]
\begin{center}
\centerline{$\begin{array}{@{}l@{}l@{}}
\textnormal{\textsf{\textbf{a}}}& \textnormal{\textsf{\textbf{b}}}\\
\multirow{4}{*}{\includegraphics[width=0.54\textwidth,valign=t]{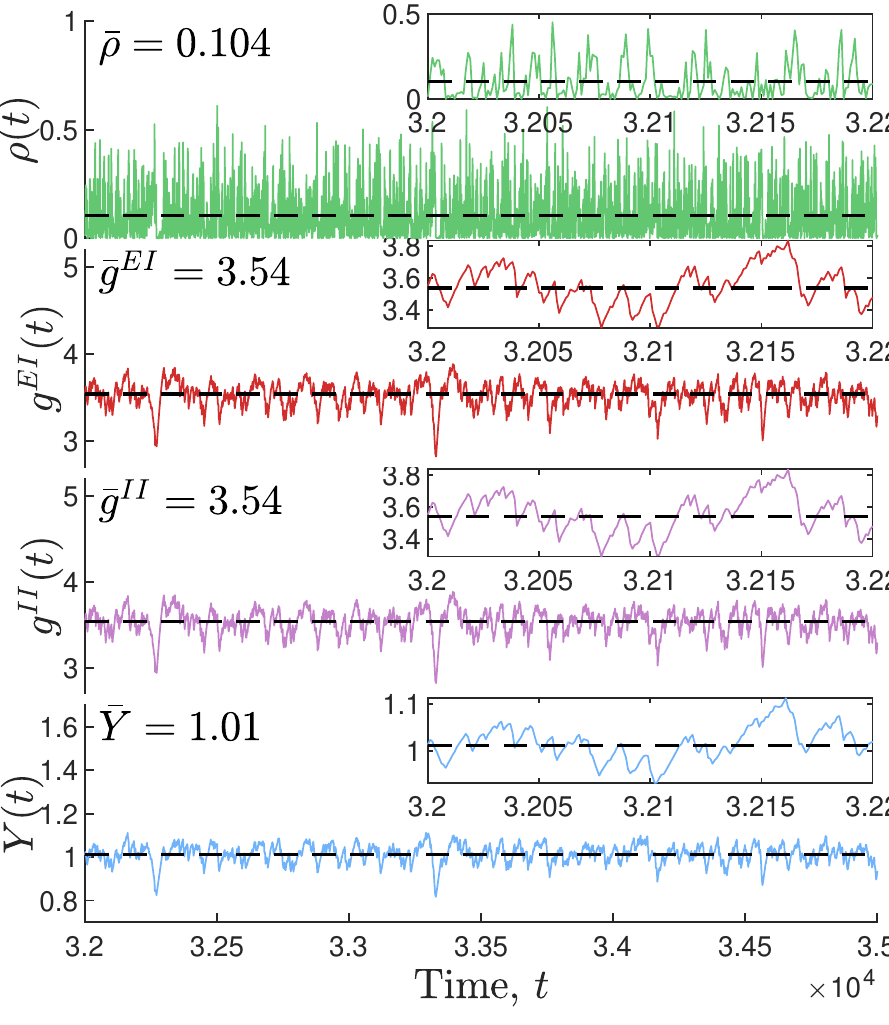}} & \includegraphics[width=0.38\textwidth,valign=t]{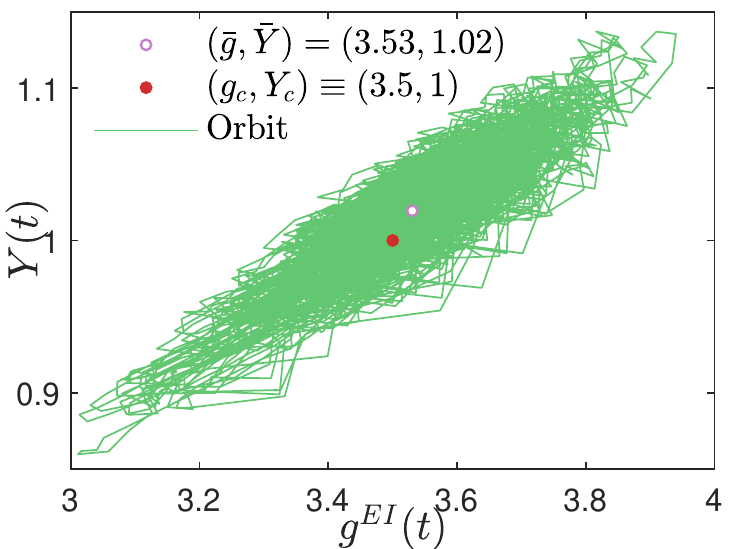}\\
& \textnormal{\textsf{\textbf{c}}}\\
& \includegraphics[width=0.38\textwidth,valign=t]{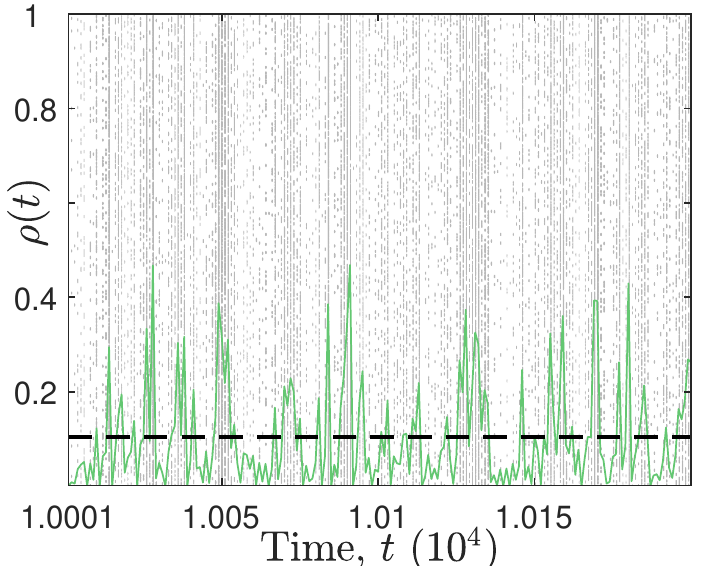}
\end{array}$}
\caption{\label{FIG10}{\bf Self-organization towards the balanced
critical point.} Parameters: $\tau_W=\tau_\theta=100$, $A = 73.5$,
$u_W=u_\theta=0.1$, $\Gamma=1$ and $J = 10$. 
{\bf a,} Time series for $\rho[t]$, $g^{EI}[t]=\avg{W_{ij}^{EI}[t]}/J$,
$g^{II}[t] = \avg{W_{ij}^{II}[t]}/J$ and $Y[t] = I/\avg{\theta_i[t]}$.
The $\rho[t]$ large peaks are dragon king events.
{\bf a (inset),} Details with a smaller time scale.
{\bf b,} Self-organization trajectories in the $g$ \textit{vs.} $Y$ plane.
The system hovers around the critical balanced point
of the static model, $g_c=3.5$ and $Y_c=1$, where neuronal avalanches
occur. The time-series average (red dot) is 
$\bar{g}^{II/EI}=3.53(2)$, $\bar{Y}=1.02(2)$.
{\bf c,} Raster plot for $1000$ neurons of the data in panel \textbf{a}.
Visually, the activity resembles the AI state, but avalanches occur and vertical
bands correspond to dragon kings (very large) avalanches.}
\end{center}
\end{figure}

In the long-term,
the proposed dynamics generates adaptation 
relative to the input level of the network
around the mean activity, given by
$\bar{g}^{II} = \overline{\avg{g_{ij}^{II}[t]}}$,
$\bar{g}^{EI} = \overline{\avg{g_{ij}^{EI}[t]}}$, and
$\bar{Y} = \overline{\avg{Y_i[t]}}$,
where the top bar denotes a long-time average and
the brackets $\avg{.}$ denote an average over the
individual activity of the neurons.
We simulated a network of $N=10^3$ neurons over a long
time ($10^4$ms=$10$s) with static parameters $\Gamma=1$,
$J=10$, $\mu=0$, $p=0.8$, and
dynamic parameters $\tau_W=\tau_\theta=100$, $A=73.5$,
$u_W=u_\theta=0.1$. In this regime,
the balance point of the static model is $g_c=3.5$ and $Y_c=1$.
In contrast, the activity of the dynamic model had averages
$\bar{g}^{II}=\bar{g}^{EI}=3.53(2)$ and $\bar{Y}=1.01(2)$,
where the digit in parentheses corresponds to the error associated to
the last digit (see Fig.~\ref{FIG10}a).

The system self-organized very close to the critical balanced
point of the static model (Fig.~\ref{FIG10}b).
However, note that $\bar{g}^{II/EI}$ and $\bar{Y}$ are slightly deviated
towards the AI region (\textit{i.e.} $\bar{g}^{II/EI}\gtrsim g_c$
and $\bar{Y}\gtrsim Y_c$, compare Fig.~\ref{FIG10}b with the
phase diagram in Fig.~\ref{PD}a). This is confirmed the
visual inspection of the network activity (Fig.~\ref{FIG10}c),
showing a global desynchronized state with local
irregular firing pattern.

\section*{Discussion}

We presented a model of an E/I network that has a balanced condition
between excitation and inhibition when $g=g_c$. The balanced point,
$g_c$, may be shifted away from the previously theoretical value $g_c\approx4$
due to the soft threshold of the stochastic neurons, keeping
the same fraction of excitatory to inhibitory neurons
in the network ($p=0.8$ for excitatory neurons,
similar to the average fraction of glutamate-activated synapses in the
brain~\cite{Somogyi1998}).

Other authors studied E/I network models and also found 
continuous phase transitions happening at points resembling
criticality~\cite{Poil2012,Lombardi2012,Lombardi2017,DallaPorta2019},
and also connecting criticality to a synchronization
phase transition~\cite{DiSanto2018}.
However, all of these models have limitations that
are naturally solved in our proposed framework.
The model by Poil~\textit{et al.}~\cite{Poil2012}
does not seem to present true criticality, since its exponents vary widely
according to the models' parameters~\cite{DallaPorta2019},
and they do not follow the proper scaling law for critical
systems~\cite{Dickman1999,Girardi2016b,Girardi2018,DallaPorta2019}.
Even though Lombardi~\textit{et al.}~\cite{Lombardi2012} also
modeled homeostatic mechanisms, their avalanche sizes do not correspond closely
to experiments neither to DP, and the avalanche distributions
follow power laws only for very small inhibition~\cite{Lombardi2017}.
Di~Santo~\textit{et al.}~\cite{DiSanto2018} studied a phenomenological model
of a spiking network and found AI and critical states,
but they do not consider E/I populations
independently, neither they derive the rate equations from a microscopic
model~\cite{DiSanto2018}.

In contrast, our model: a) generalizes the balance point
from the standard Brunel model~\cite{Brunel2000}, showing that it pertains
to the expected DP universality class of SOC systems~\cite{Dickman1999};
    b) displays the power-law scaling for avalanche sizes and duration on the
balanced critical point matching experiments~\cite{Beggs2003};
    c) contains all the four synchronicity states predicted by
Brunel~\cite{Brunel2000} when external input is considered;
    d) self-organizes towards the critical point via the adaptation of
inhibitory synapses~\cite{Deneve2016} and firing rates~\cite{Benda2003},
both of which are known to occur in the brain;
    e) the self-organized point displays quasi-critical avalanches
and a microscopic dynamics that resembles, through visual inspection, AI activity.
Thus, we unify in a single framework all observed the features from critical
neuronal network states with all the features of balanced networks,
giving rise to a self-organized (quasi-)critical balanced network.

The levels of the adaption mechanisms are controlled by their respective
time scales and fraction of potentiation.
The two required dynamics give the system a good deal of freedom in
the display of macroscopic activity. For instance, low levels of
firing rate adaptation could drive the system towards either
a switching between different synchronicity-regularity states for $Y>1$
(say, switching between SR and AR, or AR and AI, or even AI and SI),
or a self-organized bistable~\cite{diSanto2016} dynamics due to the discontinuous
transition for $Y<1$. On the other hand, low levels of inhibition adaptation 
could generate long periods of high activity intertwined with periods of low
activity. Therefore, our model could explain how drugs that modulate these
mechanisms can change the global activity of the considered system; and it could
also explain why different regions of the brain display different
activity patterns, as discussed in~\cite{Brunel2000} and~\cite{Deneve2016}
and in references therein, due to different levels of adaptability.
Nevertheless, if both mechanisms are strong, the system will be driven
towards the balanced critical point, although the effective distance
of the dynamics to the true static critical point is controlled
by the time scales of the dynamics via gross
tuning~\cite{Costa2015,Brochini2016,Costa2017,Kinouchi2019}:
the larger the time scales, the less fluctuations around
the underlying critical point are observed.

Our network is topologically regular (full graph), the
synaptic delays are fixed (1ms) and homogeneous, and
the refractory period of 1ms comes from the definition of
the neurons' dynamics. In fact, all the parameters of our model are homogeneous.
Thus, the SR, AR, AI and SI states are purely derived from the intrinsic
stochasticity of the firing of the neurons modeled by the $\Phi(V)$ function.
This function is an abstraction of all the possible noise sources happening
on the neuron's membrane. Although it is implied that the stochastic processes
on the membrane of the neurons have to occur in a time scale smaller
than the refractory period and synaptic delays of the system.

Even though we study a full graph and Brunel studies a sparse
random network~\cite{Brunel2000}, both models share the same type of results
because the author assumes that there is no correlations in the system
other than those induced by a global firing rate~\cite{Brunel2000}.
Then, his results are also in the mean-field limit due to the
high dimensionality of the network, and the critical exponents we found also apply
to his model in the balanced point. In fact, the universality of critical
phenomena guarantees that our results apply to
any network with high dimensionality that keeps the same
collective behavior, the symmetries, and the conservation laws
of our model~\cite{Odor2004}.

The simulation of large systems with the self-organizing mechanism
is limited mainly by memory due to the number of variables
that need to be kept for every time step of the
dynamics (of the order of $TN^2$, where $T$ is the total number of time
steps in the simulation). However, our results serve as proof of concept,
since increasing $N$ is known to enhance the convergence to the
critical (and balanced) point~\cite{Costa2015,Costa2017}.
Other limitation is the purely stochastic description of the spikes. Nevertheless,
if the time scales and dynamical features of the neurons and synapses are obeyed,
our results are robust due to universality~\cite{Odor2004}.

We have unified the balanced networks framework
and the neuronal avalanches SOC framework into a single formalism.
Even though the model needs deeper exploration, the general scenario is clear:
standard balanced networks are a special case of the absorbing state
SOC models in DP universality class (those with large $\Gamma J$),
and SOC models in DP are a particular case of standard balanced networks
(those with field $Y=1$). And homeostatic mechanisms in the system's inhibition
and firing rate adaptation drives the network towards a quasi-critical
that displays power-law avalanches and AI-like firing pattern.

\bibliographystyle{unsrt}

\section*{Acknowledgements}
This article was produced as part of the activities of
FAPESP Research, Innovation and Dissemination Center
for Neuromathematics (Grant No. 2013/07699-0, S. Paulo
Research Foundation). 
We acknowledge financial support
from CNPq, FACEPE, and Center for Natural and
Artificial Information Processing Systems (CNAIPS)-USP.
L.B. thanks FAPESP (Grant No. 2016/24676-1). A.A.C. thanks FAPESP
(Grants No. 2016/00430-3 and No. 2016/20945-8). M.G.-S. thanks FAPESP
(Grant No. 2018/09150-9).

\section*{Author contributions}
M.G.-S., A.A.C. and T.T.A.C. performed the simulations. O.K., L.B. and M.G.-S. made
the analytic calculations. M.G.-S. and O.K. wrote the manuscript.

\section*{Competing interests}
The authors declare no competing financial and non-financial interests.



\section*{Methods}

\subsection{Mean-field approximation}

The mean-field approximation is exact
for our complete graph network.
From the definition of the firing function, $\Phi(V)$, the model may have two types of
stationary states: the active state, such that $\rho_E = \rho_I \equiv \rho^* > 0$, and
the quiescent Q state, $\rho_E = \rho_I \equiv \rho^0 = 0$. At instant $t+1$, the
active population is simply 
given by the integral over $V$ of $\Phi(V)P_t^{E/I}(V)$
-- recalling that $\Phi(V)$ is the conditional 
probability of firing given $V$:
\begin{equation}
\rho_{E/I}[t+1] = \int_\theta^\infty \Phi(V) \:P_t^{E/I}(V) \: dV
\end{equation}
This relation is sufficient for deriving the fixed point equation for
$\rho$ when $\mu=0$.

The evolution of firing rates can also be described in terms of the neuronal
{\it firing ages} in the case $\mu>0$~\cite{Brochini2016,Kinouchi2019}.
This is possible because the reset of the potential causes a population of neurons
that fire together to also evolve together until they fire again.
We call $U_k^E[t]$ the potential of a certain population of excitatory neurons
that fired $k$ time steps before time $t$ and $\eta^{E}_k[t]$ the proportion of
such neurons with respect to the excitatory population. Then, the proportion of
firing excitatory neurons evolves as:
\begin{equation}
\rho^E[t+1]= \sum_{k=0}^{\infty} \Phi(U_k^E[t]) \: \eta_k^E[t]\:. \label{rhosumE}
\end{equation}
A neuron with firing age $k$ at time $t$ can, at time $t+1$, either fire with
probability $\Phi(U_k^E[t])$ or become part of the population with firing age $k+1$ of
proportion $\eta^{E}_{k+1}[t+1]$. In the latter case, potentials and proportions
evolve as:
\begin{eqnarray}
\eta^E_{k+1}[t+1] &=& \left( 1 - \Phi(U^E_{k}[t])\right) \eta^E_{k}[t] \label{reqetaE}\\
U^E_{k+1}[t+1] &=& \mu U^E_{k}[t] + I[t]  + p  W^{EE} \rho_E[t] - q  W^{EI} \rho_I[t] \label{reqUE} \:,
\end{eqnarray}
Where $U_0^E[t]=0$ for any $t$ (since the reset potential is zero).
By writing equivalent relations for the inhibitory
population, we obtain recurrence equations that describe the evolution of the whole
system after a reasonably short transient that guarantees that every neuron has
fired (or has had zero potential) at least once, allowing us to perform numerical
studies of the system in the mean field framework.

\subsection{The absorbing quiescent state}
\label{Q}

Let us determine the stability boundary of
the Q ($\rho^0$) phase. 
In the stationary state of equation~\eqref{Volts}, 
the distribution of voltages has a single Dirac peak.
When $V<\theta$, we have the inactive state reducing equation~\eqref{Vs} to:
\begin{equation}
V_i^{E/I}[t+1]  =  \mu V_i^{E/I}[t] + I \:,
\label{Volts}
\end{equation}
which has the stationary solution $V[t+1]=V[t]\equiv V_1$ given by
\begin{eqnarray}
\label{V1}
V_1 &=& \dfrac{I}{1-\mu} \:,\\
\label{pV}
P_{\infty}(V) &=&  \delta\left(V - V_ 1 \right) \:.
\end{eqnarray}
Since the equations~\eqref{Volts} and~\eqref{V1}
hold for both $E$ or $I$ neurons,
we dropped the $E/I$ superscripts.
Neurons start to fire when $V_1 > \theta$,
so the stability boundary line for the $Q$ phase, given by 
$V_1 = \theta$, corresponds to the dashed line
$Y_0 = I_0/\theta = 1 -\mu$ in Fig.~\ref{rhoxW}.

\subsection{The zero leakage case}

For $\mu=0$, the distribution of potentials $P_{\infty}(V)$ presents only two Dirac
peaks. The system in $V=0$ (the reset) has a population
$\rho_{E/I}[t]$ and the other, in $V=V_1^{E/I}$,
has a population $(1-\rho_{E/I}[t])$:
\begin{equation}
\label{pVact}
P_t^{E/I}(V) = \rho_{E/I}[t]
\delta\left(V\right)+\left(1-
\rho_{E/I}[t]\right)\,\delta
(V - V_1^{E/I}[t]) \:,
\end{equation}
where the  $V_1^{E/I}[t]$ are now given by:
\begin{eqnarray}
V_1^E[t] &=& I + p  W^{EE} \rho_E[t] - q  W^{EI} \rho_I[t]  \:, \nonumber\\
V_1^I[t] &=& I + p  W^{IE} \rho_E[t] - q  W^{II} \rho_I[t] \:.
\label{Vstat}
\end{eqnarray}

The firing densities $\rho_E[t]$ and $\rho_I[t]$ evolve as:
\begin{eqnarray}
\rho_E[t+1] &=& \int \Phi(V) \:P_t^E(V) \: dV \nonumber \\
& =& \left(1-\rho_E[t]\right) \left[ \Gamma \left(V^E_1 
 - \theta \right) 
 \Theta(V^E_1 - \theta) \:\Theta(V_S -V^E_1) 
 + \Theta(V_1^E - V_S) \right] \:,\nonumber \\
\rho_I[t+1] &=& \int \Phi(V) \: P_t^I(V) \: dV \nonumber\\
& =& \left(1-\rho_I[t]\right) \left[ \Gamma \left(V^I_1 
 - \theta \right) 
 \Theta(V^I_1 - \theta) \:\Theta(V_S -V^I_1)  
 + \Theta(V_1^I - V_S) \right] \:. \label{rhos2}
\end{eqnarray}

For large input current $I$, we
may have $V_1^{E/I} > V_S$ and we have:
\begin{equation}\label{cycle}
    \rho_{E/I}[t+1] = \left(1-\rho_{E/I}[t]\right) 
    \Theta(V_1^{E/I}[t] - V_S) = \left(1-\rho_{E/I}[t]\right) \:.
\end{equation}
This admits marginally stable cycle-2 solutions
(\textit{i.e.} they have no basins of attraction)~\cite{Brochini2016}. 
Equation~\eqref{cycle} has an average $\avg{\rho[t]} = 1/2$.
This fixed point corresponds to a synchronized regular (SR)
state~\cite{Brunel2000}.

When $V_1^{E/I} < V_S$, we have:
\begin{equation}
\rho_{E/I}[t+1] = \Gamma \left(V^{E/I}_1[t] - \theta \right)
\left(1-\rho_{E/I}[t]\right)  
\Theta(V^{E/I}_1[t] - \theta)  \:,
\end{equation}
For Brunel's~\cite{Brunel2000} Model A
($W^{EE} = W^{IE} = J$, $W^{II} = W^{EI} =gJ$), and using the weighted 
synaptic strength $W = pJ-qgJ$, we get:
\begin{equation}
\rho_{E/I}[t+1] = \Gamma \left(W \rho_{E/I}[t] + h \right) 
\left(1-\rho_{E/I}[t]\right) 
\Theta(W \rho_{E/I}[t] + h)\:,
\label{rhos}
\end{equation}
where we defined the field (suprathreshold current) $h = I - \theta$.

\subsubsection{The bistable phase}
\label{bistablemu0}

The bistable phase is composed
of a stable branch H ($\rho^+$), an unstable branch 
I ($\rho^-$) and the quiescent branch Q ($\rho^0$).
In the phase diagram, the bistable phase is separated
from a phase where the quiescent state is unique
through a fold bifurcation.
To determine this phase boundary, we return to
the solution for the $\rho^\pm$ fixed points:
\begin{eqnarray}
\rho^\pm & = & \dfrac{1 - g\gamma - (Y - 1)/(pJ) - 
1/(p\Gamma J) \pm \sqrt{\Delta}}{2 (1 - g\gamma)}\:, \\
\Delta & \equiv & (1 - g\gamma - (Y-1)/(pJ) - 
1/(p\Gamma J))^2 + 4 (1-g\gamma) (Y-1)/(pJ)
\nonumber \:.
\end{eqnarray}

We locate the transition line $g_1(Y)$
where the two branches appear by using the condition
$\Delta=0$. We get:
\begin{equation}
g_1 = \dfrac{p}{q} - \dfrac{1}{qJ\Gamma}\left[ 1
+ \sqrt{\Gamma (1-Y)}\right]^2  \:. \label{rhoc}
\end{equation}
This enters the phase diagram $(g,Y)$ in Fig.~\ref{rhoxg} as a first order
transition line. That means that there is no H or L states
for $g>g_1$ and $Y<1-\mu$, only the quiescent Q state,
which is fully compatible with Brunel's phase diagram~\cite{Brunel2000}.

The first order transition line can also be written as:
\begin{eqnarray}
Y_1(g,\Gamma,J) = 1 - \dfrac{1}{\Gamma} \left[
\sqrt{p\Gamma J (1 - g\gamma)} - 1 \right]^2 \:,
\end{eqnarray}
such that we recover the diagram of Brunel~\cite{Brunel2000}, with an almost 
vertical line $Y_1$ crossing $Y=1$ at
$g_1=g_c$ for large $\Gamma J$ (see Fig.~\ref{rhoxg}c).

The corresponding first order transition lines in the SOC notation are:
\begin{eqnarray}
\Gamma_1(W,h) &=&
\left[\dfrac{1}{\sqrt{W}}-\dfrac{1}{\sqrt{-2h}} \right]^2\:,\\
h_1(\Gamma,W) &=& -\dfrac{1}{\Gamma} \left[\sqrt{\Gamma W} -1\right]^2   \:,\\
W_1(\Gamma,h) &=& \dfrac{1}{\Gamma} \left[ 1
+ \sqrt{ - \Gamma h } 
\right]^2
 \:.
\end{eqnarray}
The last relation is fully compatible with equation~\eqref{rhoc},
which can be read as $W_1 =  qJ (p/q - g_1)$.
At the bifurcation point, the first order
transition step is given by:
\begin{equation}
\Delta\rho = \rho^+-\rho^0 = \dfrac{\sqrt{- \Gamma h}}{1 +\sqrt{- \Gamma h}} =
\dfrac{\sqrt{\Gamma (1-Y)}}{1 + \sqrt{\Gamma (1-Y)}}\:,
\end{equation}
which is consistent with the continuous phase
transition in the limit $h \rightarrow 0$ (see Fig.~\ref{rhoxg}a). 

\subsubsection{Transition between regular spiking states}
\label{SR}

The asynchronous regular ($\rho^+$ or AR) phase turns into
a cycle-2 (synchronous regular, SR) if its activity achieves the
value $\rho = 1/2$ as the excitation is increased.
This occurs because the neurons have a refractory period of one time step, and the
maximum firing rate of a single neuron is
$X = \{ \ldots, 1,0,1,0,1,\ldots\}$, that is, 
$\avg{X} = 1/2$.

We can obtain the transition line $W_{SR}(h)$ [or $g_{SR}(Y)$]
where the cycle-2 appear. Inserting the condition
$\rho_{SR} = 1/2$ in equations~\eqref{FOT} and \eqref{FOT2}
for $\rho^+$, we get:
\begin{eqnarray}
    W_{SR} &=& \frac{2}{\Gamma} - 2h\:,\\ 
    g_{SR}  & = & \frac{p}{q} - \frac{2}{q\Gamma J}+\frac{2h}{qJ}
    = g_c - \frac{1}{q\Gamma J} +\frac{ 2 (Y-1) }{q J} \:.
\end{eqnarray}
The result for $W_{SR}$ is coherent with previous
results for cycle-2 obtained in~\cite{Brochini2016}.
If $\Gamma J \gg 1$, the SR phase occurs for $g < g_c \approx 4$, 
independently of $Y$, which is similar to
Brunel's phase diagram~\cite{Brunel2000}.
For moderate $\Gamma J$, the transition line is
\begin{equation}
Y_{SR}(g) =  1 +\frac{1}{\Gamma}- \frac{p J}{2} \left(1-g\gamma \right)  \:,
\end{equation}
where $\gamma = q/p$ (see Fig.~\ref{rhoxg}d).

\subsubsection{Transition between irregular spiking states}
\label{SI}

Synchronous irregular (SI) appears as inhibition is increased
from an asynchronous irregular (AI) phase
through a flip bifurcation with small amplitude.
For a map of the form $\rho[t+1]=F(\rho[t])$,
the bifurcation condition is given by:
\begin{equation}
     \left. \frac{d F}{d \rho} \right|_{\rho^+} = -1 \:.
\end{equation}
where $\rho^+$ is the fixed point given by equation~\eqref{FOT}.
Simplifying the terms in the last equation, we obtain:
\begin{eqnarray}
    Y_F &=& 1 - \frac{1}{\Gamma} - W + 
    \frac{2}{\Gamma}\sqrt{1 + \Gamma W}  \:,\\
       & = & 1 - \frac{1}{\Gamma}+ pJ(g\gamma-1) +\frac{2}{\Gamma}\sqrt{1-\Gamma 
       pJ(g\gamma-1)} \:.
\end{eqnarray}
These bifurcations lines are shown in Fig.~\ref{rhoxg}d.

For large $g>g_c$, the iteration of the $\rho[t]$ map returns
$0,\Gamma (Y-1), 0, \Gamma (Y-1), \ldots$; this is a
solution of equation~\eqref{rhos}. 
This solution is also cycle-2, but its $\rho[t]<0$ part is cut off
by the step function.
To obtain it, we impose $\rho[t+2]=\rho[t]=0$.
This implies that:
\begin{eqnarray}
\rho[t+2] &=& (1-\rho[t+1]) \Gamma (h + W \rho[t+1]) \:,\\
&=& (1-\Gamma h)\Gamma (h+W \Gamma h) = 0 \:.
\end{eqnarray}
Solving this equation, we find that this cycle appears for $g>g_0$,
with $g_0$ given by
\begin{eqnarray}
\Gamma W_0 = -1 \:, \label{W0} \\
g_0 = \frac{p}{q} + \frac{1}{q \Gamma J} = 
g_c +\frac{2}{q \Gamma J}.  \label{g0}
\end{eqnarray}

\subsubsection{Low activity state with respect to the external current}

We can also examine the behavior for fixed $g > g_c$ 
as a function of input $Y$. We obtain an almost linear behavior
similar to that found by Brunel~\cite{Brunel2000}, 
see Fig.~\ref{rhoxg}b.
A first order expansion for large $J$ 
in equation~\eqref{FOT2} gives:
\begin{equation}
    \rho(Y)  \approx    \frac{Y-1 }{Jp (g \gamma -1) + (Y-1) + 
    1/\Gamma} \:\Theta(Y-1)  
    \approx  \frac{Y-1}{Jp (g \gamma -1)} 
    \: \Theta(Y-1)\:. \label{rhoY2}
\end{equation}
Equation~\eqref{rhoY2} has the correct maximum limit
$\rho \rightarrow 1$ when $Y \rightarrow \infty$.
For moderate values of $Y$, equation~\eqref{rhoY2} 
presents the linear behavior found in equation~(24)
of Brunel~\cite{Brunel2000}.

Due to leakage,
the curves of Brunel~\cite{Brunel2000}
start at $Y=1-\mu$ and have some curvature, 
in contrast to our linear 
$\rho \propto (Y-1)\: \Theta(Y-1) $ behavior
(since we used $\mu = 0$ in Fig.~\ref{rhoxg}b). 
But it is possible to obtain
curves that start at $Y = 1-\mu$ with curvature
similar to Brunel~\cite{Brunel2000}, using the solution
for $\mu>0$ below.

\subsection{Active state for nonzero leakage}

The stationary firing rates for general $\mu$ are valid in two limits:
either close to the critical region or close to the saturating regime,
i.e. when the potential of the first peak in the distribution $P_t(V)$
approaches $V_S=1/\Gamma+\theta$.

When $\mu>0$, the active phases must be described by the recurrence equations
for both excitatory and inhibitory populations. Using the simplified weight,
we can rewrite the recurrence equations~\eqref{rhosumE}, \eqref{reqetaE}
and~\eqref{reqUE} in the stationary state as:
\begin{eqnarray}
\rho&=&\sum_{k=1}^\infty\eta_k\Phi(U_k)\label{rhosum}\:, \\
\eta_k&=&\eta_{k-1}(1-\Phi(U_{k-1}))\label{reqeta} \:,\\
U_k&=&\mu U_{k-1} + W\rho+I\label{reqU}= (W\rho+I)\sum_{n=0}^{k-1}\mu^n  \\
&=& (W\rho+I)\dfrac{1-\mu^{k}}{1-\mu} \label{totUk} \:,
\end{eqnarray}
with  $U_0=0$ because we consider the resting potential to be equal to the reset
potential. Substituting equations \eqref{reqU} and \eqref{reqeta} in
\eqref{rhosum} and changing indexes yields:
\[
\rho=\sum_{k=0}^\infty\eta_k(1-\Phi(U_k))\Phi(\mu U_k+W\rho+I).
\]

Close to the critical point, where the stationary potentials belong to the
linear part of the firing function, \textit{i.e.} $\theta<U_1<U_\infty<V_S$,
where $U_\infty$ is the limit potential of ``infinite firing age''
$U_\infty=\frac{W\rho+I}{1-\mu}$. In this case:
\[
\rho= \sum_{k=0}^{\infty} \eta_k(1-\Phi(U_k)) \Gamma(\mu U_k + W \rho +I -\theta)\:. 
\]
By using the relation in equation~\eqref{rhosum} and the normalization
$\sum_{k=0}^\infty \eta^k=1$ we obtain:
\[
\rho = - \Gamma W \rho^2 + \rho(\mu + \Gamma W - \Gamma(I-\theta)) + \Gamma \tilde{h} - \Gamma \mu \sum_{k=0}^\infty \eta_k\Phi(U_k)U_k \:,
\]
where $\tilde{h}=I-(1-\mu)\theta$. Using equation~\eqref{totUk} to describe
$U_k$ only outside of the $\Phi$ function argument in the sum of the last term,
leads to:
\begin{equation}
\dfrac{\Gamma W}{1-\mu} \rho^2 + \rho (1-\mu + \dfrac{\Gamma \tilde{h}}{1-\mu} - \Gamma W) - \Gamma \tilde{h}  - \dfrac{\mu (W \rho +I)}{1-\mu}\sum_{k=0}^{\infty} \eta_k \Phi(U_k) \mu^k=0 \:,
\end{equation}
that holds when all stationary potentials lie in the linear region of the $\Phi$
function, i.e., when $U_1>\theta$ and $U_{\infty}<V_S$. Note that in the case
$\mu=0$, the above equation becomes equation~\eqref{eqrhomu0}.  

When the system is close to the critical value, we can consider the last term
to be negligible for $\mu<1$, because $\Phi(U_1)$ goes to zero as $U_1$
approaches $\theta$. In this case, we can use: 
\begin{equation}
\dfrac{\Gamma W}{1-\mu} \rho^2 + \rho (1-\mu + \dfrac{\Gamma \tilde{h}}{1-\mu} - \Gamma W) - \Gamma \tilde{h} =0\:.
\label{rhomuprox}
\end{equation}

When $\tilde{h}=0$, there is a second order phase transition because the
solutions to equation~\eqref{rhomuprox} are $\rho^-=\rho^*=0$ and 
\begin{equation}
\rho=\dfrac{W-W_c}{W}(1-\mu)\:, \label{rhomupos}
\end{equation}
where 
\begin{equation}
W_c=\dfrac{(1-\mu)}{\Gamma}\:.\label{wcmupos}
\end{equation}
This relation can also be rewritten as:
\begin{equation}
\rho=\dfrac{g_c - g}{p/q-g}(1-\mu)\:,\:\:\:\:\:\:\:\:
g_c=p/q - \dfrac{1-\mu}{ q \Gamma J} \:,
\end{equation}
in good approximation close to the critical point ($g\lesssim g_c$).

As $\rho$ approaches $1/2$, when $U_1 \geq V_S$, the $\mu>0$ case is equivalent
to the $\mu=0$ case. When $U_1<V_S$ but $U_2>V_S$, the stationary potentials
distribution has three peaks. We can then solve equation~\eqref{rhosum} with the
normalization $\eta_2=1-\eta_0 - \eta_1= 1-2\rho$.  Using this constraint in
equation~\eqref{reqeta} we obtain an equation that, interestingly, does not
depend on $\mu$:
\begin{equation}
\Gamma W \rho^2 + \rho(\Gamma h -3) +1 = 0 \:,  
\end{equation}
that provides an exact solution for the firing rate for the case $\mu>0$: 
\begin{equation}
    \rho^\pm = \frac{3-\Gamma h\pm
    \sqrt{(3-\Gamma h)^2 - 4 \Gamma W}}{2 \Gamma W} \:,
\end{equation}
that converges to $\rho^*=1/2$ as $U_1 \rightarrow V_S$. 

\end{document}